\documentclass[aps,twocolumn,showpacs,superscriptaddress,groupedaddress]{revtex4}  
\usepackage{graphicx}  
\usepackage{dcolumn}   
\usepackage{bm}        
\usepackage{amssymb}   
\usepackage{amsmath}
\usepackage{color}
\usepackage{blindtext}

\def\be{\begin{equation}}
\def\ee{\end{equation}}
\def\bea{\begin{eqnarray}}
\def\eea{\end{eqnarray}}

\begin{document}

\widetext


\title{Tetraquarks, Pentaquarks and Dibaryons in the  large $N$ QCD}
\author{Luciano Maiani}
\email{Luciano.Maiani@cern.ch}
\author{Veronica Riquer}
\email{veronica.riquer@cern.ch}
\affiliation{T. D. Lee Institute, Shanghai Jiao Tong   University, Shanghai, 200240, China}
\affiliation{Dipartimento di Fisica and INFN,  Sapienza  Universit\`a di Roma,\\ Piazzale Aldo Moro 2, I-00185 Roma, Italy}
\author{Wei Wang}
\email{wei.wang@sjtu.edu.cn}
\affiliation{ INPAC, SKLPPC, MOE Key Laboratory for Particle Physics, School of Physics and Astronomy, Shanghai Jiao Tong University, Shanghai  200240, China}
 
\date{\today}

\begin{abstract}
We study  the multiquark hadrons in large $N$ QCD under the 't-Hooft limit, extending Witten's picture of the baryons. We  explore the decay widths of tetraquarks, pentaquarks and dibaryons.  Based on the   decay behaviors, we point out in the $N\to \infty$ limit   decay widths of tetraquarks stay constant, while  those of pentaquarks and dibaryons above certain thresholds can diverge. In the large $N$ limit, we find that the ground states of the three spectroscopic series are stable or narrow and that the excited states of pentaquarks and dibaryons  above the indicated thresholds are not observables. We compare our results with those obtained in  previous  large $N$ generalizations of tetraquarks.
\end{abstract}

\pacs{14.40.Rt,12.39.-x,12.40.Yx}
\maketitle
 
\section{Introduction}

In three colour QCD,   multiquark hadrons beyond the classical $q\bar q$ and $qqq$ configurations  
may be anticipated to  exist. This expectation  stands on a long-time observation that colour antisymmetric diquarks transform as a colour ${\bar{\bf 3}}$ representation, the same as antiquarks. Thus, starting from a colour singlet hadron, one or more substitutions
\be
\bar q \to [q^\prime q^{\prime\prime}] \label{multisubs}
\ee
applied to an antiquark (and/or the corresponding substitutions for a quark) generate new, more complex, colour singlet configurations (for earlier literature, see e.g.~\cite{Jaffe:2003sg} and references therein, for recent reviews on  exotic hadrons, see e.g.~\cite{Ali:2017jda, Esposito:2016noz,  Chen:2016qju, Guo:2017jvc,Lebed:2016hpi,Olsen:2017bmm} and references therein). 
Explicitly, ignoring the  flavor structure for simplicity of notation, one has 
\begin{align}
\begin{array}{c}
 q \bar q   \\ 
  ({\rm Meson})      
\end{array} 
&\to \left\{
\begin{array}{c}
 qqq   \;\;\;
  ({\rm Baryon})   \\ 
  \bar q\bar q\bar q    \;\;\;  
  ({\rm Antibaryon})    
\end{array} \right.
\to \begin{array}{c}
 [qq][\bar q\bar q]   \\ 
  ({\rm Tetraquark})      
\end{array}  \nonumber\\
&\to \begin{array}{c}
 [qq][qq]\bar q   \\ 
  ({\rm Pentaquark})      
\end{array} 
\to \begin{array}{c}
 [qq][qq][qq]   \\ 
  ({\rm Dibaryon})      
\end{array}. 
\label{eq:replace}
 \end{align}

However, 
the extension of these concepts of  multiquarks to general $N$ is not unique. 
Taking tetraquarks as the examples,   one can construct  the diquark-antidiquark structure  in   general  $N$  according to:
 \be
 T^{(2q)}= [q^a q^{\prime b}-q^b q^{\prime a}]\times \bar q_a \bar q^\prime_b,~(a,b=1,\dots, N). \label{tetraN1}
 \ee
In this scheme, tetraquarks  manifest themselves as poles in the correlation functions of four-quark operators~\cite{Coleman:1985,Weinberg:2013cfa,Knecht:2013yqa,Lebed:2013aka,Esposito:2014rxa,Cohen:2014tga,Lucha:2017mof,Maiani:2018pef}. 
Meanwhile a formulation {\it \`a la} Witten~\cite{Witten:1979kh} for tetraquarks, the  baryonium scheme, has been explored by Rossi and Veneziano~\cite{Rossi:1977cy,Rossi:2016szw}
\begin{align}
T&= \epsilon_{a a_1  \dots a_{N-1}}\big(q^{[a_1}\dots q^{a_{N-1}]}\big) \epsilon^{a b_1  \dots b_{N-1}}\big({\bar q}_{[b_1}\dots {\bar q}_{b_{N-1}]}\big) \nonumber \\
&\equiv \left(q^{[\dots} q^{A]} \right)~\left({\bar q}_{[\dots} {\bar q_{B]}}\right)\times \epsilon_{aA} ~ \epsilon^{aB},\label{tetraN}
\end{align}
where brackets indicate antisymmetrization of the indices, $A$ and $B$ are sets of $N-1$ antisymmetrized indices and sum from $1$ to $N$ over repeated indices is understood.

\begin{figure}[htb!]
 \begin{center}
\includegraphics[width=6truecm]{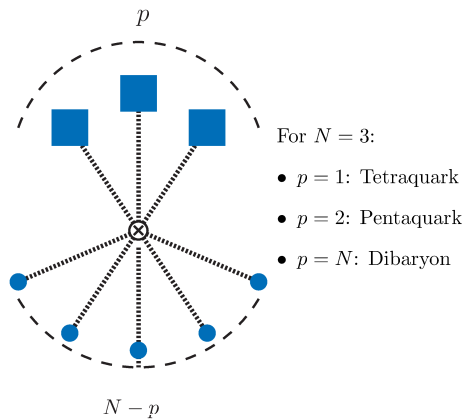} 
\end{center} \caption{\footnotesize  Multiquark hadrons generated by the substitution (\ref{multisubs}) in $N$ colours QCD. Full boxes indicate the generalised diquarks in (\ref{tetraN}) and full dots the antiquarks; dotted lines are QCD strings, the crossed circle at the center represents the $N$ dimensional $\epsilon$ symbol that joins the $N$ representations $\bar{\bm N}$ into an $SU(N)$ singlet. Hadrons resulting for $N=3$ are indicated. For brevity, the same names are used in the text for the configurations with the same value of $p$ in $N$ colours QCD.
} 
 \label{multiq}
\end{figure}

The antisymmetric combination of $N-1$ quark fields -- the generalised diquark -- transforms according to the antiquark representation $\bar {\bf N}$ and can replace any antiquark in the totally antisymmetric antibaryon. Thus, the construction  of more structures  can be performed, as in Fig.~\ref{multiq}, the first substitution  giving the tetraquark (\ref{tetraN}). A second substitution gives the large $N$ extension of the $N=3$ pentaquark~\cite{Aaij:2015tga,Maiani:2015vwa}:
\be 
P=\left(q^{[\dots} q^{A]} \right) \left(q^{[\dots} q^{B]}\right) {\bar q}_{[\dots} {\bar q_{C]}} \times \epsilon_{aA} ~ \epsilon_{bB} \times ~ \epsilon^{abC}, 
\label{pentaN}
\ee
with $C$ a set of $N-2$ antisymmetrized indices. Continuing in this way, we end with the generalised dibaryon~\cite{Jaffe:1976yi,Maiani:2015iaa}:
\bea
&&D=\left(q^{[\dots} q^{A]} \right) \dots \left(q^{[\dots} q^{B]}\right)\times   \epsilon_{aA} \dots \epsilon_{bB} \times ~ \epsilon^{a\dots b }\label{dibaN}.  
\eea
which binds $N$ generalised diquarks. 

Assigning the  baryon number  $1/N$ to one quark such that $B($baryon$)=1$,  the baryon numbers of tetraquarks, pentaquarks and dibaryons are: $0, 1, N-1$, respectively.

In a world of two colours, the above structures disappear: $N=2$ QCD is made only of mesons, $q\bar q$, ``baryons", $qq$, and   molecules thereof. 
The new  spectroscopic series start to appear at $N=3$, (our world!) and  can be  extended to $N$  colours. 

In the following, we will analyse the decay widths of the new  structures that appear at $N=3$, Eqs.~(\ref{tetraN}, \ref{pentaN}, \ref{dibaN}), dropping the qualification ``generalised" for brevity. We  will  work in large $N$ QCD under the 't-Hooft limit~\cite{Witten:1979kh}:
\be
g_s^2=\frac{\lambda^2}{N}, ~N\to \infty\label{nlarge}
\ee
where $g_s$ is the  strong  coupling  constant and $\lambda$, the 't-Hooft coupling,  is fixed.
In this limit,  some decays of generalised tetraquark in Witten's picture have been considered in Refs.~\cite{Liu:2007tj,Cohen:2014vta}.

As we will show, we find that decay amplitudes from the ground states may diverge at large $N$. However such decays are generally  forbidden by phase space and the divergent amplitudes do not affect the observability of such particles. At $N=\infty$,  ground states of multiquark hadrons are narrow or stable, particularly in the case of the dibaryon.

Results for the decay amplitudes of excited states are summarised in Tab.~\ref{sumtab}. 
For tetraquarks, we find decay amplitudes of the excited states 
that stay constant or decrease with $N$, thus confirming that the corresponding hadrons are observables. 
For excited pentaquarks and dibaryons, the de-excitation amplitudes into the ground state plus one meson are limited. However, at $N=\infty$, there are modes which have divergent amplitudes, namely $P^* \to B+B+\bar B$, $D^*\to NB+\bar B$ and $D^*\to(N-1)B$. Taken literally, these results would imply sharp thresholds at $2B+\bar B$ and $(N-1)B$ respectively, below which we expect observable pentaquarks and dibaryons, and above which we expect large, unobservable widths: a situation similar with  charmonia above and below the open charm meson-antimeson threshold.


The rest  of this paper is organized as follows. In Sect.~\ref{sec:bardec} we briefly review meson-baryon couplings in Witten's scheme to warm up and establish the notations. We consider the decays of generalised  tetraquarks in Sect.~\ref{tetradec}, and extend to the decays of generalised  pentaquarks and dibaryons in Sects.~\ref{pentadec}, \ref{dibadec}.  Discussion and  comparison  with previous large $N$ analyses 
are contained in Sect.~\ref{disc}.

\begin{table}
\begin{center}
\caption{\footnotesize Decay amplitudes for the decay of multiquark hadrons from states with only one quark excited, $A^*$. Amplitudes are normalised so that $|A^*|^2=\Gamma$, thus including a $1/\sqrt{N!}$ factor from the phase space of ${\cal O}(N)$ indistinguishable particles, see text. Entries in the Table give the leading $N$ dependence. The first two columns refer to decays  obtained from Fig.~\ref{multiq} by cutting one or more QCD strings with a $q\bar q$ pair, the third column to the decay of an excited into the ground state by one meson emission. For multi-meson or multi-baryon decay amplitudes, polynomial in $N$ appearing as prefactors of the exponentials are omitted. When giving the results for dibaryons, the diquarks inside are treated, at large $N$, as quasi classical particles.} \label{sumtab}
\begin{tabular}{||c|c|c|c|c||}
 \hline  
$T^*\to$ & {\footnotesize $B+ \bar B$} & &  {\footnotesize $T+$ Meson} & {\footnotesize Mesons}\\ \hline
$A^*\propto$ &{\footnotesize $N^{0}$}   &  &{\footnotesize $N^{0}$}&{\footnotesize $ < e^{-\frac{N}{2}}$} \\ \hline
\hline
$P^*\to$ & {\footnotesize $B+ T$} &  {\footnotesize$B+B+\bar B$} & {\footnotesize $P+$Meson}&  {\footnotesize $B+$Mesons}  \\ \hline
 $A^*\propto$ &  {\footnotesize$N^ {0}$}  &  {\footnotesize $N^{1/2}$}& {\footnotesize $N^{0}$} &  {\footnotesize $ < e^{-\frac{N}{2}}$}  \\
 \hline
\hline
$D^*\to$ && {\footnotesize $NB+\bar B$} &    {\footnotesize $D+$ Meson} &  {\footnotesize $(N-1)B$}  \\ 
\hline
$A^*\propto$ & &  {\footnotesize $> e^{+\frac{N}{2}\log N}$} &    {\footnotesize $N^{0}$}& {\footnotesize $> e^{+\frac{N}{2}\log N}$} \\ \hline
\hline
\end{tabular}
\end{center} 
\end{table}
\section {Meson-baryon couplings}
\label{sec:bardec}

As suggested by Witten, baryons  become very simple  in the limit $N\to \infty$: quarks move independently from each other  in an effective, Hartree-Fock, potential  which is $N$ independent. In the ground state, all quarks  are in the same wave-function, $\phi_0(x)$, with an $N$ independent baryon radius. 

For general states, once colour antisymmetry is guaranteed, quarks behave like a set of bosons, distributed in single particle excited states, $\phi_r$. The state is determined by the occupation numbers $n_r$, with
\be
M^*=N M_0 +\sum_r  n_r \epsilon_r,~\sum_rn_r =N,\label{excmass}
\ee
where $M_0$ and $\epsilon_r$ the excitation energy of $\phi_r$, are $N$-independent~\cite{Witten:1979kh}. We shall restrict to states with excitation energy that stays fixed when $N\to \infty$. For illustration,  we focus on:  (i) the ground state with $n_0=N$ and all other occupation numbers vanishing and (ii) the single quark excited state with $n_0=N-1$ and  $n_r=1$ for some $r$.

The normalised wave function of the above baryon states can be written as
\bea
&&{\rm {\bf ground~state}:}\nonumber \\
&&\Psi^{1~2~\dots N}_0(x_1,x_2,\dots N)=\epsilon^{1 2\dots N}\phi_0(x_1)\times  \nonumber \\
&&\times  \phi_0(x_2)\dots  \phi_0(x_N)\label{grndstat}\\
&&{\rm  {\bf excited~state}:},\nonumber \\
&&\Psi^{1~2~\dots N}_r(x_1,x_2,\dots N)=\epsilon^{1 2\dots N}\times \frac{1}{\sqrt{N}}  \nonumber \\
&&\times [{\bm  \phi_r}(x_1)\phi_0(x_2)\dots  \phi_0(x_N)+\phi_0(x_1){\bm  \phi_r}(x_2)\dots  \phi_0(x_N)\nonumber \\
&& +\dots +\phi_0(x_1)\phi_0(x_2)\dots {\bm  \phi_r}(x_N)]. 
\label{exctdstat}
\eea
Here $1,2,\dots,N$ are colour indices. Since the  flavour and spin indices are omitted,  $\Psi$ are antisymmetric in colour (to make a colour singlet) and fully symmetric in   coordinate space.

\begin{figure}[htb!]
 \begin{center}
\includegraphics[width=4truecm]{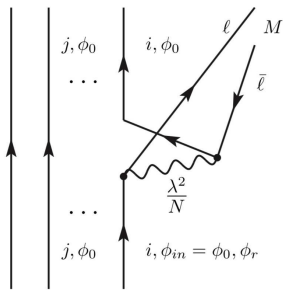}
\end{center} \caption{Feynman diagrams for meson-baryon trilinear coupling in the large $N$ limit. An excited quark emits a gluon, turning into a quark-antiquark pair with the factor $\lambda^2/N$. 
} 
 \label{barexc}
\end{figure}

The meson-baryon trilinear coupling is represented in Fig.~\ref{barexc}. The initial quark wave function is indicated by $\phi_{in}$, the final quark is in the ground state $\phi_0$. With $\phi_{in}=\phi_0$ or $\phi_r$, we obtain the ground state meson-baryon coupling, e.g. $g_{N\bar N \pi}$, or the transition amplitude of the excited state, e.g. $A(\Delta \to N \pi)$. 

The basic transition occurs via one gluon exchange: $q^i(\phi_{in})\to q^i (\phi_0)+q^\ell +{\bar q}^\ell$. Projecting over the colour singlet meson state
\be
M=\frac{1}{\sqrt{N}} \sum_\ell \bar q_\ell q^\ell,\label{mesonstate}
\ee
one finds the effective operator for the baryon to baryon transition
\be
{\it O}=\frac{\lambda^2}{N} \sqrt{N}~\bar q  {\cal O}(x) q. 
\ee
In the above, the ${\cal O}(x)$ is a $N$-independent operator acting on the single quark wave functions $\phi(x)$ in (\ref{grndstat}) and (\ref{exctdstat}) and connecting $\phi_{in}$ to $\phi_0$.

The transition operator applied to  (\ref{grndstat}) gives $N$ equal terms and we obtain~\cite{Witten:1979kh}:
\be
g_{B\bar B M}\sim \frac{\lambda^2}{N} \sqrt{N} ~N\propto \sqrt{N}.\label{barcoup}
\ee

When applied to (\ref{exctdstat}), the transition operator has to operate on $\phi_r$ only, to obtain a non-vanishing result when the scalar product with (\ref{grndstat}) is taken. We obtain $N$ equal factors, divided by the normalisation $\sqrt{N}$, so that
\be
A(B^* \to B+M)\sim \frac{\lambda^2}{N} \sqrt{N}~\frac{N}{ \sqrt{N}} \propto N^0\label{bardec}.
\ee



It is interesting to notice that the tree-level meson-baryon low energy scattering amplitude obtained from (\ref{barcoup}) is ${\cal O}(1)$ since the baryon's  propagator brings in a factor $N^{-1}$ due to the baryon's mass~\cite{Witten:1979kh}.

\section {Tetraquark decays}
\label{tetradec}

  \begin{figure}[htb!]
 \begin{center}
\includegraphics[width=6truecm]{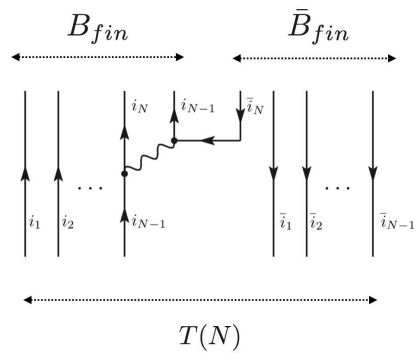}
\end{center} \caption{Feynman diagrams for the tetraquark decays into a pair of baryons, $T\to B+ \bar B$. 
} 
 \label{tetrawit1}
\end{figure}

We first consider   the natural decay of the tetraquark, the baryonium mode:
\be
T\to B+ \bar B, \label{tbbar}
\ee
which is depicted in  Fig.~\ref{tetrawit1}. The Feynman diagram is basically the same of the previous Section, but we  must be careful with the normalizations of initial and final states, which differ by $N$ dependent factors.  

We normalise the mesons $M$, baryons $B$ and tetraquarks $T$ as 
\be
\langle 0|M M^\dagger |0\rangle=\langle0|BB^\dagger |0\rangle=\langle0|T T^\dagger|0\rangle=1. 
\ee
For mesons the above normalization is realized in (\ref{mesonstate}) while  for baryons, we have 
\be
B=q^1 q^2\cdots q^N, \bar B=\bar q_1\bar q_2 \cdots \bar q_N\label{barN}, 
\ee
which is consistent with  Eq.~(\ref{grndstat}).

For tetraquarks, we use the wave function:
\be
T=\frac{1}{\sqrt{N}} B_a\bar B^a\label{normatetra},
\ee
where $B_a=\frac{\partial}{\partial q^a }B$ is the operator $B$ with $q^a$ suppressed, similarly for $\bar B^a$ with $\bar q_a$ and the sum over $a=1,\cdots N$ is understood. 

Due to the anticommutation properties  of quark fields, there might be a minus sign in the definition of $B_a$, depending from the position of $q^a$ in (\ref{barN}). However, one has the identity
\be
q^a B_a=q^a\frac{\partial}{\partial q^a} B= B, {\rm for~any}~a
\label{signfix}
\ee 
since $q^a\partial/\partial q^a$ is a bosonic operator.

Consider in Fig.~\ref{tetrawit1} the case where quark lines in the initial diquark correspond to $B_1$. Then $i_{N}=1$ and 
the pair created by the gluon interaction provides the missing $q^1$ and $\bar q_1$ to the product $B_1\bar B^1$. 
We need to add all diagrams where the gluon is emitted by the other quark lines, which gives a factor $N-1$. Finally, summing over $a$ gives a factor of $N$, since all the terms of the sum are equal to the one with $a=1$.

Given various anticommutation signs, one can suspect that terms arising from different values of $a$ come with different signs, but one can show that it is not so. The gluon interaction from the $i^{th}$ quark produces the substitution:
\be
q^i \to q^i q^a\bar q_a 
\label{g-rad}
\ee 
Since  $q^a\bar q_a$ is a bosonic operator, the result from the term $B_a \bar B^a$ can be written as
\bea
&& B_a \bar B^a ~({\rm no~sum~on}~a) \to B_a \cdot q^a\bar q_a \cdot \bar B^a\nonumber \\
&&= (-1)^{(N-1)}q^a \frac{\partial}{\partial q^a}B\cdot \bar q_a \frac{\partial}{\partial \bar q_a} \bar B\nonumber \\
&&=(-1)^{(N-1)}\cdot B \bar B, 
\eea
which is independent on $a$.

Adding the diagrams in which  gluon emission occurs from an antiquark line gives   a factor two to the final result.

In conclusion, putting all together, we find the scaling: 
\be
 A(T\to B+ \bar B)=\frac{1}{\sqrt{N}}~\frac{\lambda^2}{N}~(N-1)~N  \propto{\sqrt{N}}. 
 \label{grdstbbar}
 \ee

The decay in Eq.~(\ref{tbbar}) is unlikely to occur for  the ground state tetraquark, since the $B\bar B$ state has one pair of constituent quarks compared to  the initial state, and there is not enough phase space.

Instead, the decay in $B\bar B$ will occur for an excited (radial or orbital) state $T^*$, where the excitation energy $\epsilon_r-\epsilon_0$ can be used to create the additional pair that will  transform the tetraquark into $B\bar B$. A consequence is that only the excited quark emits the gluon in Fig.~\ref{tetrawit1} and we loose a factor of $\sqrt{N-1}$ as in Sect.~\ref{sec:bardec}:
\be
 A(T^*\to B+ \bar B) \propto N^0 \label{tetrastar1}.     
 \ee
The analogy with Eqs.~(\ref{bardec}) and (\ref{barcoup}) is evident.

Taking~the amplitudes in (\ref{grdstbbar}) and~(\ref{tetrastar1}) as proportional to the couplings of $g_{TB\bar B}$ and $g_{T^* B\bar B}$, we may estimate the large $N$ behaviour of the scattering amplitude $B+\bar B \to B+\bar B$ via tetraquark intermediate state. We find:
\bea
&& A(B+\bar B \to B+\bar B) \propto\nonumber \\
&&\propto   g_{TB\bar B}\frac{1}{M_T^2}g_{TB\bar B}+\sum_{T^*}~ g_{T^*B\bar B}\frac{1}{M_T^2}g_{T^*B\bar B}\nonumber\\
&&\propto {\cal O}(N^{-1})+ \sum_{T^*}~ {\cal O}(N^{-2}).
\label{BBbarscatt}
\eea 
since $ M_T^2\sim M_{T^*}^2\propto N^2$. We find a leading contribution of order $N^{-1}$ from the ground state  in agreement with the estimate in \cite{Rossi:2016szw}, see their Eq. (26), and a nonleading one from the excited states of order $N^{-2}$.

  \begin{figure}[htb!]
 \begin{center}
\includegraphics[width=6truecm]{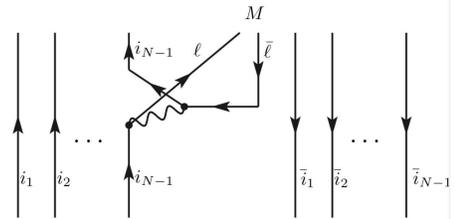}
\end{center} \caption{Feynman diagrams for the decay of an excited tetraquark to the ground  state by emitting a light meson. The diagrams with the deexcitation from the antidiquark only produce a factor of two
} 
 \label{tdexc}
\end{figure}

We  now consider  the decay of an excited tetraquark to the ground  state by emitting a meson, which is shown in Fig.~\ref{tdexc}.
There  is  a factor $(1/\sqrt{N})^2$ from the normalization of the initial and final tetraquarks, a factor $1/\sqrt{N}$
for the normalization of the meson, a factor $N$ for the number of $q^\ell \bar q_\ell$ pairs, a factor $N$ for the number of diquarks with different colours, each of which gives a factor $\sqrt{N-1}$ for the transition from an excited quark. 
Multiplying by the factor $1/N$
from the gluon interaction, we obtain
\be
A(T^* \to T+M)\propto N^0. 
\ee
The diagrams with the deexcitation from the antidiquark only produce a factor of two.

  \begin{figure}[htb!]
 \begin{center}
\includegraphics[width=5truecm]{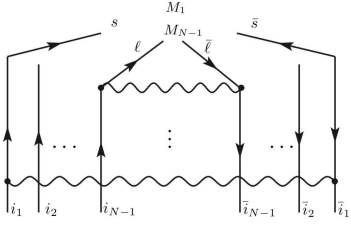}
\end{center} \caption{Feynman diagrams for the tetraquark decays into $N-1$ mesons. 
} 
 \label{tetrawit2}
\end{figure}

We continue the discussion on tetraquarks  with the decay into mesons:
\be
T\to M_1 +\cdots +M_{N-1}.
\ee
Quarks lines in the diquark are paired with the corresponding antiquark lines, to form $N-1$ quark-antiquark mesons, depicted in Fig.~\ref{tetrawit2}. 
The complex of these diagrams lead to the effective Lagrangian
\be
{\cal L}_{eff}=\frac{g(N)}{(2m)^{N-4}} M^{N-1} T \label{leffdim}
\ee
and
\be 
g(N)\propto \frac{N}{\sqrt{N}} \left(\frac{\lambda^2}{\sqrt{N}}\right)^{N-1}\propto N^{\frac{1}{2}(N-1)} 
\ee
$m$ is the quark mass and the dimensional factor, needed to makes $g(N)$ adimensional, will be cancelled by similar dimensional factors appearing in the computation of the rate see Appendix. 
We find, omitting finite powers of $N$
\bea
&&\Gamma \propto  \frac{g(N)^2}{(2m)^{2(N-4)}}\frac{(N-1)!^2}{(N-1)!}\times \nonumber \\
&& \frac{1}{2M_T}\int \frac{dk_1 dk_2 \dots dk_{N-1}}{(2\mu)^{N-1}}~ \delta(Q_T-\sum_i \frac{k_i^2}{2\mu})  \delta^{(3)}(\sum_i {\bm k}_i)\nonumber\\   
\label{ratetetra}
\eea
the $(N-1)!^2$ in the numerator comes from the contraction of the operator $M^{N-1}$ with the external meson states, the $(N-1)!$ in the denominator from the phase space of $N-1$ identical mesons. In the phase space integral, $\mu$ is the meson mass and $Q_T$ the $Q$ value of the decay, using Eq.~(\ref{excmass}):
\be
Q_T=M(T)-(N-1)\mu =(2M_0-\mu)(N-1)\label{qtetra} 
\ee
The phase space integral is discussed in Appendix and shown to give a decreasing contribution at large $N$. An {\it upper bound} to the rate  is obtained by keeping factorials only. Omitting polynomial prefactors, we find
\bea
&&\Gamma < g(N)^2 (N-1)!\sim \frac{N!}{N^N}=e^{-N};\nonumber \\
&& A(T\to M_1 +\cdots +M_{N-1})=\sqrt{\Gamma} < e^{-\frac{N}{2}}. \label{multimsexp}
\eea
Up to a polynomial in $N$, the result holds for the decay amplitude of the excited tetraquark, $A^*$, as given in Tab.~\ref{sumtab}.

Finally, as a way to check our method, we compute the amplitude for the decay of a tetraquark to a complex $T_{-2}=(N-2)q+(N-2)\bar q$, using the  diagram in Fig.~\ref{tetrawit2} restricted to the exchange of one gluon.

Following Eq.~(\ref{normatetra}), we define 
\be
B_{ab}=\frac{\partial}{\partial q^a}\frac{\partial}{\partial q^b} B=- B_{ba}. 
\label{signfix2}
\ee
with $B_{ab}$ antisymmetric for any value of $N$.
The normalized $T_{-2}$ is defined as
\begin{align}
T_{-2}=\frac{1}{\sqrt{N(N-1)/2}}   \sum_{a< b} B_{ab}~ \bar B^{ab},\;\;a,b = 1\cdots, N.
\label{normam2}
\end{align}

From Fig.~\ref{tetrawit2} we find:
\bea
&&T=\frac{1}{\sqrt{N}} \sum_a ~B_a \bar B^a  \nonumber \\
&&\to   \frac{1}{\sqrt{N}} ~\frac{\lambda^2}{N} ~\sum _{b>a} B_{ab} {\bar B}_{ab}~\sum_\ell q^\ell{\bar q}_\ell \nonumber \\
&&= \frac{1}{\sqrt{N}} ~\frac{\lambda^2}{N} \sqrt{N} M~\sqrt{\frac{N(N-1)}{2}} T_{-2}  \nonumber \\
&& \propto N^0~ M\cdot T_{-2}
\eea 
in agreement with \cite{Cohen:2014vta}.

\section {Pentaquarks}
\label{pentadec}

In analogy to Eqs.~(\ref{normatetra}) and~(\ref{normam2}), one  can  simplify the expression describing  normalized pentaquarks according to
\begin{align}
P=\frac{1}{\sqrt{N(N-1)/2}}~\sum_{a< b} B_a B_b~ \bar B^{ab},\;\;a,b = 1\cdots, N
\label{pentaN2}. 
 \end{align}

The antisymmetry in colour has to be matched to the symmetry under exchange of the two diquarks in Eqs.~(\ref{pentaN}) and (\ref{pentaN2}), which are fermions or bosons for $N=$ even or odd, respectively. 

To see how this works, we smear out the coordinates of the generalized diquarks with a function $F(x,y,z)$, according to
\bea
&&P=\frac{1}{\sqrt{2N(N-1)}}~\int dx dy dz ~F(x,y,z)\times \nonumber \\
&& \times \sum_{a\leq b}B_a(x) B_b(y)~ \bar B^{ab}(z). 
\eea
Thus, Bose or Fermi symmetry under the exchange: $x,a \leftrightarrow y,b$ implies
\be
F(y,x,z)=(-1)^N F(x,y,z). 
\label{symmN1}
\ee

For simplicity, we have assumed diquarks with equal flavour and spin distribution, in which case (\ref{symmN1}) implies diquarks in relative $P$-wave for $N$ odd. Unequal flavour and spin properties, however, allow $S$ and $P$-wave diquarks, as considered in~\cite{Maiani:2015vwa}.

A case in point is the calculation of the decay into a baryon and a tetraquark, which goes through the process (\ref{g-rad}), where the gluon is radiated from either one or the other diquark.  The process transforms the initial state (\ref{pentaN2}) according to (overall signs are ignored)
\bea
&&P\to \frac{1}{\sqrt{N(N-1)/2}} \left(\frac{\lambda^2}{N}\right) (N-1)\times \nonumber \\
&&\times\sum_{a< b} \left[\bar q_a q^a B_a B_b+(-1)^N B_a \bar q_b q^b B_b\right]\bar B^{ab}\label{pentabt},
\eea
where $(-1)^N$ is the statistical factor associated to the exchange of diquark coordinates, (\ref{symmN1}). Using Eqs.~(\ref{signfix},~\ref{signfix2}), we have 
\bea
&&\sum_{a< b} \left[\bar q_a q^a B_a B_b+(-1)^N B_a \bar q_b q^b B_b\right]\bar B^{ab}\nonumber \\
&&=-B\sum_{a< b}\left(B_b \bar B^b+B_a \bar B^a\right) \nonumber\\
&&=-B\left[\sum_{b=1}^N B_b \bar B^b (b-1)+\sum_{a=1}^N B_a \bar B^a (N-a-1)\right]\nonumber \\
&&=-(N-2)\sqrt{N} ~B\cdot T. 
\eea

One should add pair creation from the antiquarks in $B^{ab}$, with the antiquark completing the antidiquark and the quark absorbed by one diquark to make the baryon. This process gives an amplitude of the same form as (\ref{pentabt}) (dictated by colour conservation and statistics) and the same $N$ dependence, so that, 
\be
A(P\to B+ T)\propto \sqrt{N} \label{pentabar}. 
\ee

As before, decay from the ground state is forbidden by phase space while the  decay amplitude    from an excited pentaquark is reduced by a factor $\sqrt{N-1}$
\be
A(P^*\to B+ T)\propto N^0. 
\ee

For other decay modes, 
we  obtain similarly: 
\bea
&& A(P \to B+B+\bar B) = {\cal O}(N)\label{pentalead},  \\
&& A(P^* \to P+M)\propto N^0\label{pentadeexc},\\
&& A(P,~P^* \to B+{\rm Mesons})< e^{-\frac{N}{2}}. 
\eea

For decay of a pentaquark excited state, $P^*$, to the final state in (\ref{pentalead}), we have a reduction of a factor $\sqrt{N}$ in amplitude, Tab.~\ref{sumtab}, which, however, is stil divergent for $N\to\infty$.

\section{Dibaryons}\label{dibadec}
Starting from the definition, Eq.~(\ref{dibaN}), we smear the coordinates of each generalised diquark with a function $F(x_1,\cdots,x_N)$
\bea
&&D=\int dx_1\cdots dx_N~F(x_1,\cdots,x_N)\times \nonumber \\
&&\times \epsilon^{\alpha \cdots \beta}~B_\alpha(x_1)\cdots B_\beta(x_N)\label{dibasmear0}. 
\eea

For diquarks with identical quark spin and flavour composition,  $F(x_1,x_2,\cdots,x_N)$
must be  symmetric (antisymmetric) in the exchange of any two coordinates for $N=$ even (odd). 

For  $N=3$, the role of spin and flavour composition is well illustrated by the first example of a dibaryon considered in the literature, the so-called $H$-dibaryon introduced by Jaffe~\cite{Jaffe:1976yi} and based on the antisymmetric scalar diquarks of $SU(3)_{flavor}$
\bea
&&H=\epsilon^{\alpha \beta\gamma} ([ds]_0)_\alpha(x_1) ([su]_0)_\beta(x_2) ([ud]_0)_\gamma(x_3)\nonumber \\
&&=\epsilon^{\alpha \beta\gamma}\phi_{\alpha 1}(x_1)w_{\beta2} (x_2)w_{\gamma 3}(x_3)\nonumber \\
&&\sim \epsilon^{\alpha \beta\gamma}\epsilon^{ijk}~\phi_{\alpha i}(x_1)w_{\beta j} (x_2)w_{\gamma k}(x_3), 
\eea
where the subscript $0$ indicates the total spin of the diquarks, $i,j,k$ are $SU(3)_{flavor}$ indices. An $S$-wave dibaryon, fully symmetric under coordinate exchanges, is made possible by the antisymmetry in flavour of the bosonic diquarks.

Starting from (\ref{dibasmear0}), we restrict from now on to equal spin and flavour composition. We can reduce all terms multiplying the fully antisymmetric tensor to the basic permutation and write
\bea
D&=&\int dx_1\cdots dx_N~F(x_1,\cdots,x_N)\nonumber \\
&&\times B_1(x_1)\cdots B_N(x_N)\label{dibasmear}. 
\eea


We consider first the decays into many baryons, $D\to N B+\bar B$.
The decay is produced by the emission of one $q\bar q$ pair from each diquark, Eq.~(\ref{g-rad}), with each quark joining its diquark to form a baryon  and the antiquarks forming the antibaryon.   Starting from (\ref{dibasmear}) we obtain
\bea
&&D\to \left(\frac{\lambda^2}{N}\right)^N (N-1)^N \int dx_1\cdots dx_N~F(x_1,\cdots,x_N)  \nonumber \\
&&\times  [\bar q_1(0) q^1(x_1) B_1(x_1)]\cdots [\bar q_N(0) q^N(x_N) B_N(x_N)]. 
\eea
After using Eq.~(\ref{signfix}) and ignoring overall signs, we obtain
\bea
&&D\to \left(\frac{\lambda^2}{N} \right)^N  (N-1)^N \int dx_1\cdots dx_N~F(x_1,\cdots,x_N)  \nonumber \\
&&\;\;\;\; \times   \bar B(0) B(x_1)\cdots B(x_N). 
\eea
The effective Lagrangian for the  $D\to N B + \bar B$ can be constructed as: 
\bea
&&{\cal L}_{eff}=g(N) \times \left\{ \begin{array}{l} (M_0)^{-\frac{3N}{2}+\frac{3}{2}},~(N={\rm odd}) \\ (M_0)^{-N+2},~(N={\rm even}) \end{array}\right.\times \nonumber \\
 &&\times \left(B^{\dagger N}{\bar B}^\dagger D+ {\rm h.c.}\right);  \nonumber \\
&& g(N) \propto N^0. \label{leffnbplusbbar}
\eea
To compute the decay rate, we consider first the case of $N=$ odd, in particular $N=3$, where baryons are fermions. The extension to  even $N$, when F is symmetric, is obvious and it gives the same result.

Suppose that in the dibaryon, diquarks are in single diquark wave functions $w_0(x),w_1(x) \dots$. We use a different symbol for the wave functions, to distinguish the motion of diquarks,  which are quasi classical particles, with energy levels that vanish at $N=\infty$, from the motion of quarks in the diquark, which are fully quantum with energy level spacing of order ${\cal O}(\Lambda_{QCD})$.

The effective lagrangian has baryons in the same wave functions as the original diquarks, and one needs to consider the fully antisymmetric combination, represented by the so-called Slater determinant. We write explicitly the $N=3$ case
\bea
&&F(x_1,x_2,x_3~|~a,b,c)=C\times \nonumber\\
&& \times {\rm Det} \left[\begin{array}{ccc}w_a(x_1) &w_b(x_1) &w_c(x_1)\\w_a(x_2) &w_b(x_2) &\phi_c(x_2)\\ w_a(x_3) &w_b(x_3) &w_c(x_3) \end{array}\right]. 
\eea
In the above, C is a normalization constant, the determinant is the sum of $3!$ products $w_{a^\prime}(x_1)w_{b^\prime}(x_2)w_{c^\prime}(x_3)$ where $a^\prime,b^\prime,c^\prime$ is a permutation of $a,b,c$, multiplied by $+ 1$ ($- 1$) if the permutation is even (odd). The $w$ are orthogonal and these terms make an orthonormal system, so that the normalization  is $C= {1}/{\sqrt{3!}}$. 

Omitting proportionality constants independent from $N$, the transition matrix element is
\bea
&&T=\langle B(k_1), B(k_2), B(k_3), \bar B(p)|{\cal L}_{eff} | D\rangle\nonumber \\
&&=g(3)\frac{1}{\sqrt{3!}}~ 3~!\times {\tilde F}(k_1,k_2,k_3~|~a,b,c)\times f(k),
\eea
where $f(k)$ is a function of the momenta and $\tilde F$ contains the Fourier transforms of the $\phi$ s. 

The $3!$ in the numerator arises because each monomial, say $B(x_1)^\dagger B(x_2)^\dagger B(x_3)^\dagger w_a(x_1)w_b(x_2)w_c(x_3)$, when contracted with the external baryons gives rise to the full Slater determinant $\tilde F(k_1,k_2,k_3~|~a,b,c)$ with the appropriate sign.

Given the orthonormality of $\tilde F$ and passing to general $N$=odd, one has:
\bea
&& \Gamma=|A|^2\propto g(N)^2 N!^2 \frac{1}{N!} \times \nonumber \\
&& \int dk_1 dk_2 ~\dots dk_{N+1} \delta(Q_D-\sum_i \frac{k_i^2}{2M}) f(k)\label{dibarnbpbbar}
\eea
where $M$ is the nucleon mass and $Q_D$ the $Q$ value. Extending Eq.~(\ref{excmass}) to the ground state, one would obtain a negative $Q$ value, so we consider directly the excited state, where
\bea
&&Q_D=M_D^* - (N+1) M=\nonumber \\
&&=M_0^* (N-1) N - N^2 M_0 \sim (M_0^*-M_0)N^2\label{dibarq}
\eea

As shown in Appendix, the phase space integral is divergent for large $N$ so that a {\it lower bound} to the rate is obtained by taking the factorials only and we find:
\be
A(D^*\to N B+\bar B)=\sqrt{\Gamma}> e^{+\frac{N}{2}\log N}\label{grdst1}. 
\ee
as given in Tab.~\ref{sumtab}, up to polynomial prefactors. It is necessary to mention that the results are based on the treatment that the diquarks are quasi classical particles. 

\begin{figure}[htb!] 
\begin{center}
\includegraphics[width=6truecm]{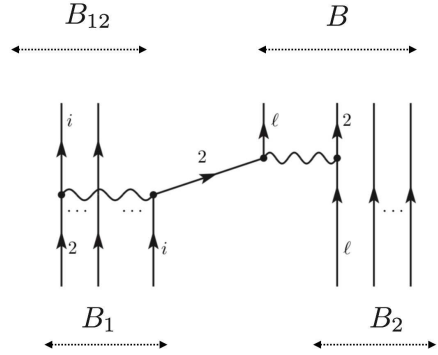}
\end{center} 
\caption{Feynman diagrams for the dibaryon decays into $N-1$ baryons, $D\to (N-1) B$.   } \label{dbdec2}
\end{figure}

We next consider $D\to (N-1) B$. The basic element is reported in Fig.~\ref{dbdec2}, which llustrates the exchange of quark $q^2$ between the generalised diquarks $B_1$ and $B_2$, completing the latter into a baryon and transforming the former in $B_{12}$. Transferring $q^3$, $q^4$, etc. to $B_3$, $B_4$ etc. one obtains a final state with $N-1$ baryons. The amplitude of Fig.~\ref{dbdec2}, taking into account the different options for quarks $i$ and $\ell$, is
\be
A(B_1 + B_2 \to B_{12} + B)\propto (\frac{\lambda^2}{N})^2 (N-2) (N-1). 
\ee
One can proceed considering the transfer of $q^3$ to $B_3$ and so on. Finally, one replaces the role of $B_1$ with $B_2$, etc.

Collecting all factors, one obtains the effective lagrangian
\bea
&&{\cal L}_{eff}=g(N)\times \left\{ \begin{array}{l}(M_0)^{-\frac{3}{2}(N-1)+3},~(N={\rm odd}) \\(M_0)^{-N+4},~(N={\rm even}) \end{array}\right.\times \nonumber \\
&&\times \left(B^{\dagger (N-1)}D + {\rm h.c.}\right), \nonumber\\
&&  g(N)\propto \frac{N!}{N^N}\sim e^{-N}, 
\eea
and, proceeding as in the previous case
\be
A(D\to (N-1)B)> \frac{N!^{3/2}}{N^N}\sim e^{+\frac{N}{2}\log N}. 
\label{grdst2}
\ee
The result holds for decays of the excited dibaryons as well, in our approximation of neglecting polynomial prefactors.

Finally, we consider $D^*\to D+$Meson. The basic process is similar with Fig.~\ref{barexc}, adapted to the excited dibaryon initial state. In the $N\to\infty$ limit, diquarks have infinite mass and can be considered as classical objects. We may assume that the excited diquark is $B_1$:
\begin{itemize}
\item a factor $\sqrt{N-1}$ for the gluon emitted by a quark in $B_1$ with the transition $w_r \to w_0$;
\item  a factor $\sqrt{N}$ for the meson;   
\item  a factor $\lambda^2/N$ for gluon exchange.
\end{itemize}
In total, we have 
\be
A(D^*\to D+M) \propto N^0 \label{dibdeexc}. 
\ee

If allowed by phase space, the results (\ref{grdst1}) and (\ref{grdst2}) would make the width of the dibaryon ground state unobservably large.  However, as observed in~\cite{Jaffe:1976yi} for the $H$-dibaryon, colour and spin-spin attraction may push the mass of the ground state below the $N$ baryons threshold as well, and make the ground state dibaryon to be stable under strong decay \footnote{an even lighter, stable $H$-dibaryon has been advocated in~\cite{Farrar:2017ysn} as the source of the astrophysical dark matter, see~\cite{Gross:2018ivp} for an updated discussion.}. A similar phenomenon has been advocated for diquarks made by two heavy quarks~\cite{Esposito:2013fma} to predict doubly heavy tetraquarks to be stable against strong decays~\cite{Karliner:2017qjm,Eichten:2017ffp,Eichten:2017ual}. 

In these scenarios, the amplitude (\ref{grdst2}) is ineffective, the ground state decays weakly and the de-excitation amplitudes of lightly excited dibaryons go to a constant value at large $N$, Eq.~(\ref{dibdeexc}).

\section {Discussions}
\label{disc}

In summary, we have extended Witten's description of baryons in large $N$ QCD~\cite{Witten:1979kh} to the multiquark hadrons generated by replacing one or more antiquarks in an antibaryon with the generalised diquark made by $N-1$ quarks in the antisymmetric colour combination. The first step reproduces the generalised tetraquark considered by Rossi and Veneziano~\cite{Rossi:2016szw}. Successive substitutions produce the large $N$ generalisation of multiquark configurations considered for $N=3$: pentaquarks~\cite{Maiani:2015vwa} and dibaryons~\cite{Jaffe:1976yi,Maiani:2015iaa}.

We have studied in the present paper the decays into conventional baryons and mesons of  ground and low-lying excited states, namely states with a finite energy difference with respect to the ground state, in the limit $N\to \infty$.

Decay amplitudes  for the ground states generally diverge with $N$. However, we have argued that the final configurations with divergent amplitudes are forbidden by phase space. In this case, we would expect ground states with narrow widths, for tetra and pentaquarks where multi-meson states  are available, or decay by weak interactions, in the case of dibaryons where also $N$ baryon states are expected to be phase-space forbidden.  

The decay amplitudes of multiquark low lying excited states are reported in Tab.~\ref{sumtab}. The first two columns refers to decays obtained from Fig.~\ref{multiq} by cutting one or more QCD strings with a $q\bar q$ pair, the third column to the decay of an excited into the ground state by meson emission. The last column refers to decays obtained by reorganising the quark-antiquark pairs of the initial state into a multimeson state or in redistribuiting the quarks of one diquark to the other diquarks, to form a set of $N-1$ baryons.

A few remarks are given in order. 
\begin{itemize}
\item {\bf excited tetraquarks}: the amplitudes for the decay of the excited states do vanish or remain constant for $N\to \infty$, therefore leading to observable states in this limit;
\item tetraquark de-excitation amplitudes and $B\bar B$ decay amplitudes are of the same order;
{  \item For $N=3$ and flavour composition $[cu][\bar c\bar u]$ the threshold for  two-baryon decay is $2M(\Lambda_c)\sim 4570$~MeV; in Ref.~\cite{Cotugno:2009ys} it is argued that $X(4660)$ is a $P$-wave tetraquark decaying predominantly into $\Lambda_c \bar \Lambda_c$ in addition to the mode into $\psi(2S) \pi \pi$~\cite{Dai:2017fwx}.}
\item tetraquark-charmonium mixing is exponentially suppressed;
\item {  excited pentaquarks and dibaryons}:~de-excitation amplitudes into the ground state and a meson remain limited for large $N$;
 \item at $N=\infty$ there are modes which give divergent amplitudes, namely $P^* \to B+B+\bar B$ and $D^*\to NB+\bar B$ or $(N-1)B$; \item taken literally, these results, imply sharp thresholds at $2B+\bar B$ and $(N-1)B$ respectively, below which we expect observable pentaquarks and dibaryons, and above which we expect large, unobservable widths: a situation similar to charmonia above and below the open charm-anticharm meson threshold.
 { \item For $N=3$ and pentaquark with flavour composition:~$[cu][ud]\bar c$, corresponding to the states observed by LHCb~\cite{Aaij:2015tga}, the threshold for ``non-observabilty" would be  $2M(\Lambda_c) + M(P)\sim~5510$~MeV, for a double charmed dibaryon with flavour $[cu][cd][ud]$ the threshold would be at: $2M(\Lambda_c)$.
 }
\end{itemize}

For tetraquarks, we agree with \cite{Cohen:2014vta} for multiquark decays and for the decay into $T_{-2}+$ Meson. 

We have added the decay into $B+\bar B$ which brings in a divergent behaviour for the ground state. 
The divergence at large $N$ is not relevant for the width and the observability of the ground state, which is below threshold for the decay, but it makes it dominant as intermediate state in elastic  $B+ \bar B$ scattering, see Eq.~(\ref{BBbarscatt}). The $1/N$ behaviour we find for the latter amplitude is in agreement with the result given in~\cite{Rossi:2016szw}. 

Also novel is the result about the de-excitation of a tetraquark to the ground state plus a meson, expected to have a constant limit for $N\to\infty$ and  to be phenomenologically important.

Finally, it is interesting to compare the results for tetraquarks with  the analysis based  on the large $N$ generalisation of diquarks given in~(\ref{tetraN1}). 
The results in~\cite{Maiani:2018pef} feature
\begin{itemize}
\item a suppressed decay amplitude of the ground state into two mesons, of order $N^{-2}$; for large $N$ this is larger that the exponentially suppressed amplitude in Tab.~\ref{sumtab}, but it takes $N\geq 6$ for the power suppression to  win over the exponential suppression (see~\cite{Maiani:2017kyi} for a further suppression of two meson decay due to the potential barrier opposing $q\bar q$ annihilation in the diquark-antidiquark picture); 
\item amplitude of order $N^{-1/2}$ for the de-excitation into the ground state by meson emission;
\item tetraquark-charmonium mixing occurs to order $N^{-3/2}$; 
\item the decay of an excited tetraquark  into $B\bar B$ cannot be computed.
\end{itemize}

 The similarities of two very different multiquark generalisations encourage us to think that they support the the $N=3$ interpretation of exotic hadrons as reflections of diquark dynamics.
 
\acknowledgements 
We thank A. Polosa for many interesting discussions, G. C. Rossi and G. Veneziano for a useful exchange on tetraquak decays, C. Liu as well as T. Cohen, F. J. Llanes-Estrada, J. R. Pelaez and J. Ruiz de Elvira for useful correspondence on their works. V. R. thanks Prof. Xiangdong Ji for hospitality at the T. D. Lee institute, where this work has been done. This work is supported in part by  National  Natural
Science Foundation of China under Grant
 No.11575110, 11735010, 11747611,  Natural  Science Foundation of Shanghai under Grant  No.~15DZ2272100.

\vskip01cm

\appendix
\section{\\ Phase space  integral at large $N$}

In connection with Eq.~(\ref{ratetetra}), we study here the behaviour of  $n$-particles phase space integral for large $n$. In the relativistic case and for large $n$ we approximate the momentum $\delta$-function as
\be
\delta^{(3)}(\sum_i {\bm k}_i)\sim \delta^{(3)}(0)\sim (2m Q)^{-\frac{3}{2}}
\ee
and consider:
\be
{\cal I}(n)= \int dk_1 dk_2 ~\dots dk_{n} ~(2m Q)^{-\frac{3}{2}}~\delta(Q-\sum_i \frac{k_i^2}{2m})
\ee
Defining
\be
K^2=\sum_i k_i^2
\ee
the integral is 
\bea
&&{\cal I}(n)= (2m Q)^{-\frac{3}{2}} \int dk_1 dk_2 ~\dots dk_n ~ \delta(K^2-\sum_i \frac{k_i^2}{2m})=\nonumber \\
&&=m(2m Q)^{-\frac{3}{2}} C(3n) K^{(3n-2)} 
\eea
where $C(3n)K^{3n-1}$ is the surphace of the hypersphere of radius $K$ in $3$n dimensions, that is
\be
C(3n)=\frac{2\pi^{\frac{3n}{2}}}{\Gamma(\frac{3n}{2})}
\ee
so that
\be 
{\cal I}=\frac{2m\pi^{\frac{3n}{2}}}{\Gamma(\frac{3n}{2})}
~(2mQ)^{\frac{3n}{2}-\frac{5}{2}}\label{intps}
\ee

\paragraph{\bf{Tetraquark to mesons.}} From~(\ref{qtetra}) we read
\be
Q_T=(2M_0-\mu)(N-1)=\delta Q_T(N-1)
\ee
$\delta Q_T$ is essentially the difference between the masses of two quarks bound in a nucleon or in the meson, $\delta Q_T <<2\mu$. 

Setting $n=N-1$ and neglecting non leading terms except in dimensionfull terms, we find  
\bea
&&{\cal I}=(2\mu)^{3N-7} (\frac{\delta Q_T}{3\mu})^{(\frac{3N}{2})}\frac{(\frac{3N}{2})^{(\frac{3N}{2})}}{\Gamma(\frac{3N}{2})}\sim \nonumber \\
&&\sim (2\mu)^{(3N-7)} (\frac{\delta Q_T}{3\mu})^{(\frac{3N}{2})}\label{tetralim}
\eea
having used Stirling's fomula. The factor $(2\mu)^{(3N-7)}$ is absorbed by corresponding factors in the formula for the rate and in the dimensions of the coupling, Eqs.~(\ref{ratetetra}) and~(\ref{leffdim}), to yield the correct dimension of the decay rate. In conclusion, 
\be
\Gamma \propto g(N)^2 e^{-N[\frac{3}{2}\log(\frac{3\mu}{\delta Q_T})+C]}
\ee
where $C$ is the logarithm of all the adimensional factors like $2\pi$s, meson to quark mass ratio etc.. For small $\delta Q_T$ 
\be
\Gamma\propto  e^{-N[1+\frac{3}{2}\log(\frac{3\mu}{\delta Q_T})]}<e^{-N}
\ee
as reported in Eq.~(\ref{multimsexp}) and Tab.~\ref{sumtab}.

\paragraph{\bf{Dibaryon to $N B+\bar B$.}} In this case, $n=N+1$ and, see Eqs.~(\ref{excmass}) and (\ref{dibarq}),
\be
Q_D \sim (M^*_0-M_0) N^2=\delta Q_D N^2
\ee
 where $\delta Q_D$ is essentially the mass difference of a quark mass in the dibaryon or in the baryon. Using~(\ref{leffnbplusbbar}), (\ref{intps}) and inserting the normalization factors for $N+1$ fermions or bosons as appropriate, one finds from Eq.~(\ref{dibarnbpbbar}), 
\bea
&&\Gamma \sim g(N)^2 N! \frac{(M_0)^2}{M_D} \left(\frac{\delta Q_D}{M_0}\right)^{\frac{3}{2}}\times \left\{\begin{array}{c} N^{3N},~({\rm N=odd}) \\ N^{2N},~({\rm N=even})\end{array}\right..\nonumber \\
\eea


\end{document}